\documentclass[prl, twocolumn, showpacs,amsmath,amssymb]{revtex4-1}

\usepackage[T1]{fontenc}                                                                                                                                                                                                   
\usepackage[latin9]{inputenc}
\usepackage{hyperref}
\usepackage{graphicx}
\usepackage{xcolor}
\usepackage{braket}
\usepackage{commath}
\usepackage{amsmath}
\usepackage{ulem}
\newcommand{\bs}[1]{{\boldsymbol{#1}}}
\newcommand{\bk}{\bs{k}}
\newcommand{\br}{\bs{r}}

\definecolor{mygreen}{RGB}{20,148,20}
 
\usepackage{filemod}

\begin{document}

\title{Quantum boomeranglike effect of wave packets in random media} 

\author{Tony Prat$^1$, Dominique Delande$^1$, Nicolas Cherroret$^1$}
\affiliation{$^1$Laboratoire Kastler Brossel, UPMC-Sorbonne Universit\'es, CNRS, ENS-PSL Research University, Coll\`ege de France; 4 Place Jussieu, 75005 Paris, France}

\begin{abstract}
We unveil an original manifestation of Anderson localization for wave packets launched with a finite average velocity: after an initial ballistic motion, the center of mass of the wave packet experiences a retroreflection and  slowly returns to its initial position, an effect that we dub "Quantum Boomerang" and describe numerically and analytically in dimension 1.
In dimension 3, we show numerically that the quantum boomerang  is a genuine signature of Anderson localization: it exists if and only if the quantum dynamics if localized. 
\end{abstract}

\maketitle

Anderson localization (AL), the absence of wave diffusion due to destructive interference in disordered potentials \cite{Anderson1958}, is ubiquitous in condensed-matter systems, 
wave physics or atom optics. This offers many experimental platforms for its characterization, as was demonstrated experimentally with light \cite{Chabanov2000, Schwartz2007} 
(see however \cite{Storzer2006, Sperling16}) or ultrasound waves \cite{Hu2008}. Recently, AL of  atomic matter waves has also 
been observed \cite{Chabe08, Billy2008, Jendrzejewski12, Manai15, Semeghini15}, as well as its many-body counterpart \cite{Schreiber15, Choi16}. 
A precious asset of atom optics experiments is to allow for direct tests of fundamental manifestations of AL, such as the time evolution of wave packets. 
In this context, a common experimental scenario for probing localization consists in preparing a spatially narrow  atomic wave packet in a trap,
then opening the trap and monitoring the time evolution of the gas \cite{Modugno10, Shapiro12}. After release, the wave packet spreads symmetrically 
around its initial position and quickly becomes localized in space. What happens, now, 
if a nonzero average velocity is additionally imprinted to the gas? 
In a classical picture, one expects the randomization of velocities due to scattering on the random potential to stop the initial ballistic motion of 
the wave packet center-of-mass (CoM) at roughly a mean free path $\ell$, and then a symmetric localization of the packet around this new central position 
due to AL. We show in this article that the evolution is in fact very different:
if the quantum dynamics is Anderson localized, after an initial ballistic motion where the 
CoM indeed increases up to $\ell$, the wave packet slowly \textit{returns to its initial position}, recovering a symmetric shape at long time. In contrast, if the quantum dynamics is diffusive, e.g. in dimension 3 above the mobility edge, the CoM evolves at long time toward a final position different from the initial one.

In this article, we thoroughly study  this phenomenon that we dub Quantum Boomerang (QB) effect.
In dimension 1, we give an exact solution. In dimension 3, we show numerically 
that the QB effect exists for strong disorder where the dynamics is Anderson localized, and is partly destroyed for disorder strengths where the long time dynamics is diffusive: the CoM evolves ballistically at short time, makes a U-turn at intermediate time, but does not completely
return to its initial position. We further show that it is a faithful signature of AL, allowing to precisely pinpoint the position of the mobility edge.

Let us start with a one-dimensional (1D) system described by the Hamiltonian $ H = -\hbar^2 \Delta /(2m) + V(x)$, where $V(x)$ is a Gaussian, 
uncorrelated random potential: $\overline{V(x)}=0$ and $\overline{V(x) V(x')} = \gamma \delta(x-x')$, where the overbar denotes averaging over disorder realizations. 
We wish to study the time evolution of  a normalized Gaussian wave packet, $\Psi_{k_0}(x) \!\propto\!\exp \left[-x^2/(2 \sigma^2)  + i k_0 x \right]$, 
to which a finite momentum $\hbar k_0\!>\!$ is imprinted. 
To simplify the discussion, we assume throughout this article a sharp initial velocity distribution, $k_0\sigma\!\gg\!1$, and weak disorder, $k_0\ell\!\gg\!1$, thereby allowing for a simple description of the wave packet
in terms of two velocity components $ \pm \hbar k_0 / m$, with energy $E_0\!=\!\hbar^2k_0^2/2m$. 

\begin{figure}[h]
 \includegraphics[width=0.9\linewidth]{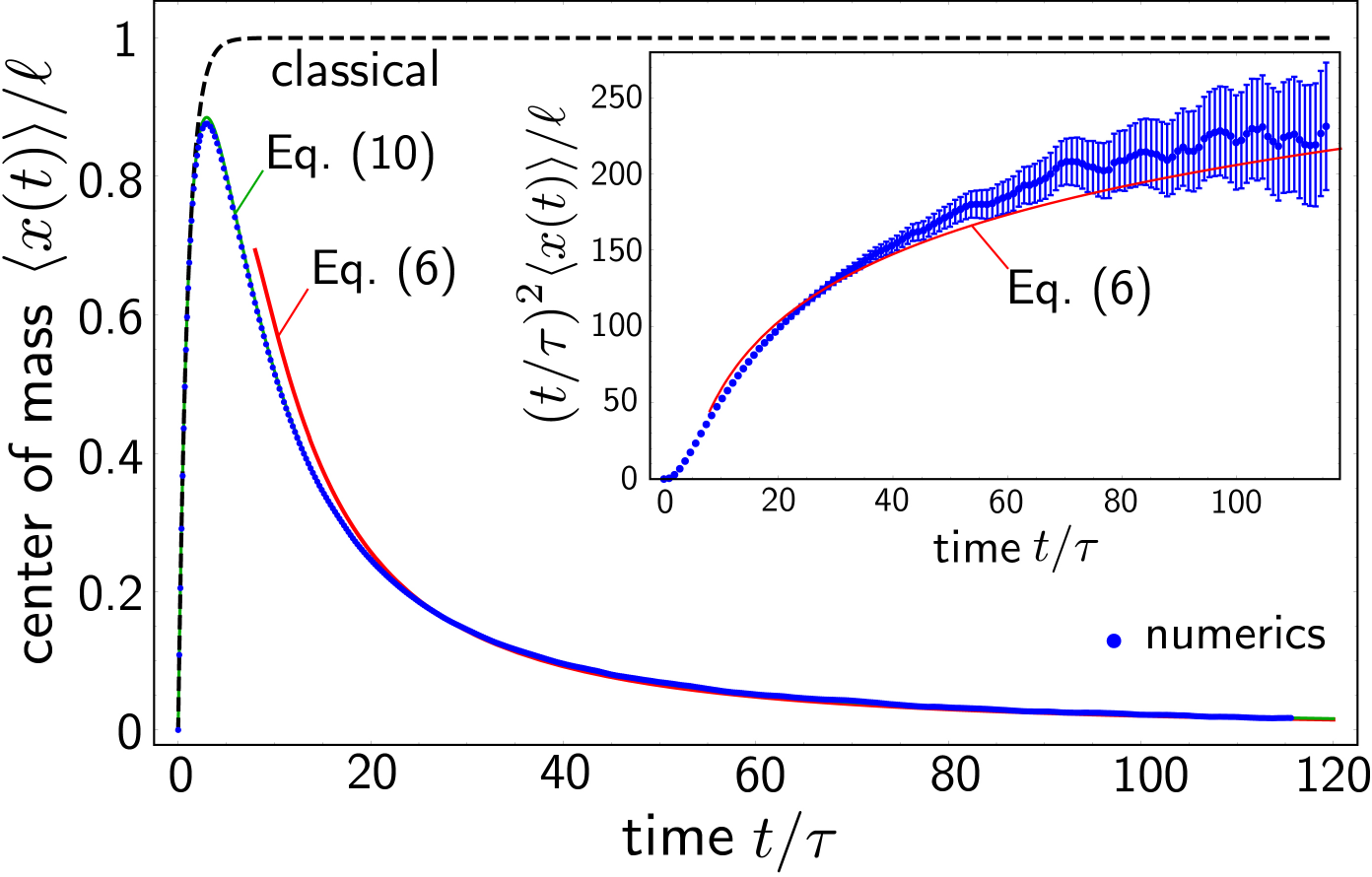}
 \caption{
 	Main plot: center of mass as a function of time. 
 Its long-time asymptotics,  Eq. \eqref{results_mean_x}, is shown as a solid red curve. The re-summation of the short-time series, Eq. \eqref{eq-Pade-x}, (solid green curve) perfectly overlaps with the numerical result (blue dots). The dashed curve is the classical result, Eq. \eqref{eq-x-classic}.
 Inset: center of mass multiplied by $(t/\tau)^2$ as a function of time.
 The asymptotic result (\ref{results_mean_x}) (red curve) is compared to the numerical prediction, displayed with its statistical error bars. 
The parameters used are given in the main text.}
 \label{fig-mean_x}
\end{figure}
The average evolution in the random potential is governed by two microscopic scales,  the scattering mean free time $\tau$ and the scattering mean free path $\ell\!=\!v_0\tau$, where $v_0\!=\!\hbar k_0 / m$. In the following, $\tau$ and $\ell$ are calculated to the leading order in 
$1/k_0\ell\!\ll\!1$, using the Born approximation at energy $E_0$ \cite{AkkermansMontambaux}.
The assumption of uncorrelated random potential is not crucial for our discussion: all the results hold as well for short-range correlated potentials, 
provided that $\ell$ and $\tau$ are replaced by the transport mean free path and time, respectively \cite{Gogolin82}.

By numerically propagating $\Psi_{k_0}(x)$, we obtain the disorder-averaged density profile $\overline{|\Psi(x,t)|^2}$, 
from which we compute the CoM $\langle x(t)\rangle\equiv \int x \overline{|\Psi(x,t)|^2} dx$. The result is shown in Fig. \ref{fig-mean_x} :
$\langle x(t)\rangle$ first increases rapidly, reaches a maximum at $t\sim\tau$ and then slowly decreases to zero. In other words, after a transient motion rightward, 
the wave packet slowly \textit{returns to its initial position} $x=0$.
For these simulations we discretize the Hamiltonian on a 1D grid 
of size $16000 \pi /k_0$, divided into 251352 grid points.
The initial wave packet width is set to $\sigma = 10/k_0$, and $\gamma=0.0058 \hbar^4k_0^3/m^2$  ($k_0\ell=\hbar^4k_0^3/2m^2\gamma \simeq  86.5 $).
The results are averaged over 45000 disorder realizations.  In the simulations, 
the evolution operator is expanded in a series of Chebyshev polynomials, as explained in \cite{Roche97,Fehske09}.
The behavior observed in Fig. \ref{fig-mean_x} is dramatically different from the classical expectation, 
which can be simply deduced from Ehrenfest theorem: $\partial_t \braket{x}_\text{class} = 
\langle p\rangle /m\!=\!\hbar k_0(n_+\!-\!n_-) /m$
where $n_\pm$ is the population of particles with momentum $\pm\hbar k_0$ ($n_+\!+\!n_-\!=\!1$). Using the classical Boltzmann equations $\partial_t n_\pm\!=\!(n_\mp\!-\!n_\pm)/(2\tau)$ 
with the initial condition 
$n_+\!=\!0$, we find
\begin{equation}
 \braket{x(t)}_\text{class}= \ell \left( 1 - e^{-t / \tau} \right).
 \label{eq-x-classic}
\end{equation}
Within the classical picture, the CoM thus quickly saturates to the mean free path $\ell$, but never experiences retroreflection.
The reason why quantum wave packets behave so differently can be understood by the following argument. 
At any time, the density distribution can be expanded over the eigenbasis $\{\epsilon_n , \ket{\phi_n} \}$ of $H$ as
\begin{align}
\begin{split}
\vert \Psi(x,t) \vert^2 = & \sum_{n,m} \braket{\phi_n \vert \Psi_{k_0}}  \braket{\Psi_{k_0} \vert \phi_m}
\\
&\times  \phi_n(x) \phi_m^*(x) e^{-i(\epsilon_n - \epsilon_m)t/\hbar}.
\end{split}
\label{eq-exp-psi-eigenstates}
\end{align}
Since eigenstates are localized, the system is constrained to a volume set by the localization length $\xi\!=\!2 \ell$. This defines a  typical mean level spacing $\Delta\!=\!1/\rho\xi$ ($\rho:$ density of states per unit volume), with a corresponding localization time $\tau_\text{loc}\!=\!2\pi\hbar/\Delta\!=\!4\tau$ beyond which the off-diagonal oscillatory terms $n\!\neq\!m$ in Eq. (\ref{eq-exp-psi-eigenstates}) vanish, leaving:
\begin{equation}
\overline{\vert \Psi (x,\infty) \vert^2} =  \sum_{n} \overline{ \vert \braket{\phi_n \vert \Psi_{k_0}}\vert^2 \vert \phi_n(x) \vert^2 }.
\label{eq-inf-time-prof-mode-exp}
\end{equation}
Due to time-reversal invariance, the $\phi_n(x)$ are real so that $ \braket{\phi_n \vert \Psi_{k_0}}\!=\!\braket{\phi_n \vert \Psi_{-k_0}}^*$:
Eq. (\ref{eq-inf-time-prof-mode-exp}) is independent of the sign of $k_0$, and thus coincides with the long-time, 
spatially symmetric density distribution that would have been obtained with an initial wave packet having a symmetric velocity distribution.
This shows that the CoM \emph{must} return to its initial position at long times, as a result of AL. Note that this conclusion is also valid in arbitrary dimension $d$ if the eigenstates are Anderson localized, with a typical mean level spacing  $\Delta\!=\!1/\rho\xi^d$, see below. 

\begin{figure}
\begin{center}
\includegraphics[scale=0.9]{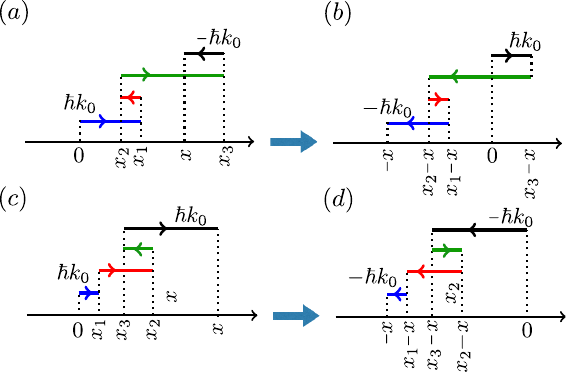}
\end{center}
 \caption{ Scattering paths contributing to the center of mass. a) A typical multiple
 		scattering path going from $x\!=\!0$ to $x,$ contributing to $\braket{x}_-$ (the path is unfolded to the top for clarity). 
 		The momentum reverses at each scattering event.  By time-reversing and translating this path by $-x$, we obtain path b), 
 		which gives an opposite contribution to $\braket{x}_-$, ensuring that $\braket{x}_-$ vanishes. c) Path contributing to $\braket{x}_+$. 
 		Its time-reversed and translated counterpart d) starts with momentum $-k_0$, not populated at $t=0$, so that $\braket{x}_+\!\neq\!0$.}
 \label{fig-diagrams}
\end{figure}
Let us now be more quantitative and analyze the CoM at finite times. For this purpose, we start by applying the Ehrenfest theorem to the mean-square displacement, 
$ \partial_t \braket{x^2}\!=\!\braket{\left[ x^2,p^2 \right]} /2 i \hbar m$, and split the particle distribution into two classes of positive and negative velocities:
$\overline{\vert \Psi(x,t) \vert^2}\!=\!n_+(x,t)\!+\!n_-(x,t)$. This leads to \cite{footnote4}
\begin{equation}
 \partial_t \braket{x^2} = 2 v_0 \braket{x}_+  - 2v_0 \braket{x}_-.
  \label{Ehrenfest_x2_v0}
\end{equation}
Here $\braket{x}_\pm=\int_{-\infty}^\infty x\ n_\pm(x,t)\ dx$, with obviously $\braket{x}=\braket{x}_++ \braket{x}_-$. 
We now consider an arbitrary path contributing to $\braket{x}_-$ [Fig. \ref{fig-diagrams}(a)]. The path starts at $x=0$ with momentum $\hbar k_0$ 
and reaches  $x$ with momentum $-\hbar k_0$ at time $t$. By time-reversing and translating this path of a distance $-x$, 
one can always find a complementary path starting with momentum $\hbar k_0$ at $x=0$ and reaching $-x$ at time $t$ [Fig. \ref{fig-diagrams}(b)]. 
Due to time-reversal and translational invariance, these two paths contribute with the same weight to $n_-(x,t)$, 
which is thus an even function of $x$, yielding $\braket{x}_-$ = 0. 
This reasoning does not apply to ${\braket{x}_+}$ since the time-reversed/translated counterpart of an arbitrary path contributing to ${\braket{x}_+}$ starts with momentum $-\hbar k_0$ which is not initially populated [see Figs. \ref{fig-diagrams}(c)-(d)]. We have thus
\begin{equation}
 \partial_t { \braket{x^2} }= 2v_0 { \braket{x} },
 \label{eq-relation_x_xx}
\end{equation}
a property that we can use to infer the long-time limit of $\braket{x}$ from $\braket{x^2}$, previously computed in \cite{Nakhmedov87}. It yields \cite{footnote5}
\begin{equation}
{\braket{x(t)}} = \ell \frac{  64  \text{ln}(t/4\tau) \tau^2}{t^2} + \mathcal{O} \left( \frac{1}{t^2} \right).
\label{results_mean_x}
\end{equation}
Eq. (\ref{results_mean_x}) is shown in Fig. \ref{fig-mean_x}
and matches well the exact numerical prediction at long time. The inset of Fig. \ref{fig-mean_x} also confirms the presence of the logarithmic term
in Eq.~(\ref{results_mean_x}).

One can go one step further and exploit Eq. (\ref{eq-relation_x_xx}) to compute $\braket{x(t)}$ at \textit{any time}.
For this purpose, we use the Berezinskii diagrammatic technique \cite{Berezinsky73} which, combined with Eq. (\ref{eq-relation_x_xx}), gives \cite{footnote5}
 \begin{equation}
   \braket{x(t)} = \int \frac{\text{d} \omega}{2 \pi} e^{-i \omega t } \left[ - \frac{2 \ell}{i \omega} \sum_{m=0}^{\infty} P^1_m(\omega) Q^1_m(\omega) \right],
 \end{equation}
where $P^1_m(\omega)=s \Gamma(m+1) [ \Psi( m+1,2;- s) - (m+1) \Psi( m+2,2;- s) ]$,
with $s= 4 i \omega \tau$, $\Gamma$ the Gamma function and $\Psi$ the confluent hypergeometric function of the second kind. The $Q^1_m(\omega)$ are solutions of
\begin{eqnarray}
 &&[4 i \tau (m + 1/2) \omega-(m+1)^2-m^2] Q^1_m(\omega) \nonumber \\
 &&+ (m+1)^2 Q^1_{m+1}(\omega) + m^2 Q^1_{m-1}(\omega) + P^1_m(\omega) = 0.
 \label{eq-Q1-alpha}
\end{eqnarray}

At short time, one can solve these equations with the expansion $Q^1_m(\omega) = \sum_{n=0}^{+\infty} q_{m,n}/(i \omega)^{n}$.
To compute the $q_{m,n}$, we first notice that $q_{m,i} = 0 \text{ if } i \leq m$, which follows from the large-frequency expansion of $P_m^1(\omega)$ (which has no terms $1/\omega^i$ with $i < m$). We use this result to expand Eqs. (\ref{eq-Q1-alpha}) order by order in $1/\omega$ and reduce them to a triangular system.
This method provides us with the coefficients $\chi_n$ of the expansion  $\braket{x(t)}=\ell \sum_n\chi_n(t/\tau)^n$ at arbitrary order.
We find for instance
 \begin{equation}
\braket{x(t)}= \ell \left[ \frac{t}{\tau} - \frac{t^2}{2 \tau^2} + \frac{t^3}{6 \tau^3} - \frac{3 t^4}{64 \tau^4} \right] + \mathcal{O} \left( t^5 \right).
\label{results_mean_x_short-t}
\end{equation}
The method cannot be directly used to estimate $\braket{x(t)}$ at any time because the series has a 
finite convergence radius, estimated at $4\tau$ from the first 100 terms.
Nevertheless, the observed exponential decay of the $\chi_n$ makes this series a good candidate for a Pad\'e resummation. 
The knowledge of the long-time limit \eqref{results_mean_x} suggests to express the CoM at any time under the form
\begin{equation}
 \braket{x(t)} = \ell \frac{\text{ln}(1+t/ 4 \tau) \tau^2}{t^2} \lim_{n\to\infty} R_{n}(t),
 \label{eq-Pade-x}
\end{equation}
where $R_{n}(t)$ is a diagonal Pad\'e approximant of order $n$, deduced from the $\chi_n$ coefficients \cite{Baker75}. 
In practice, $R_{n}(t)$ converges quickly, and an excellent approximation of $\langle x(t)\rangle$ for times up to $120 \tau$ is obtained with $n=7$. 
This is demonstrated by the solid green curve in Fig. \ref{fig-mean_x}, which perfectly coincides with the numerical results.
\begin{figure}
	\begin{center}
		\includegraphics[width=0.94\linewidth]{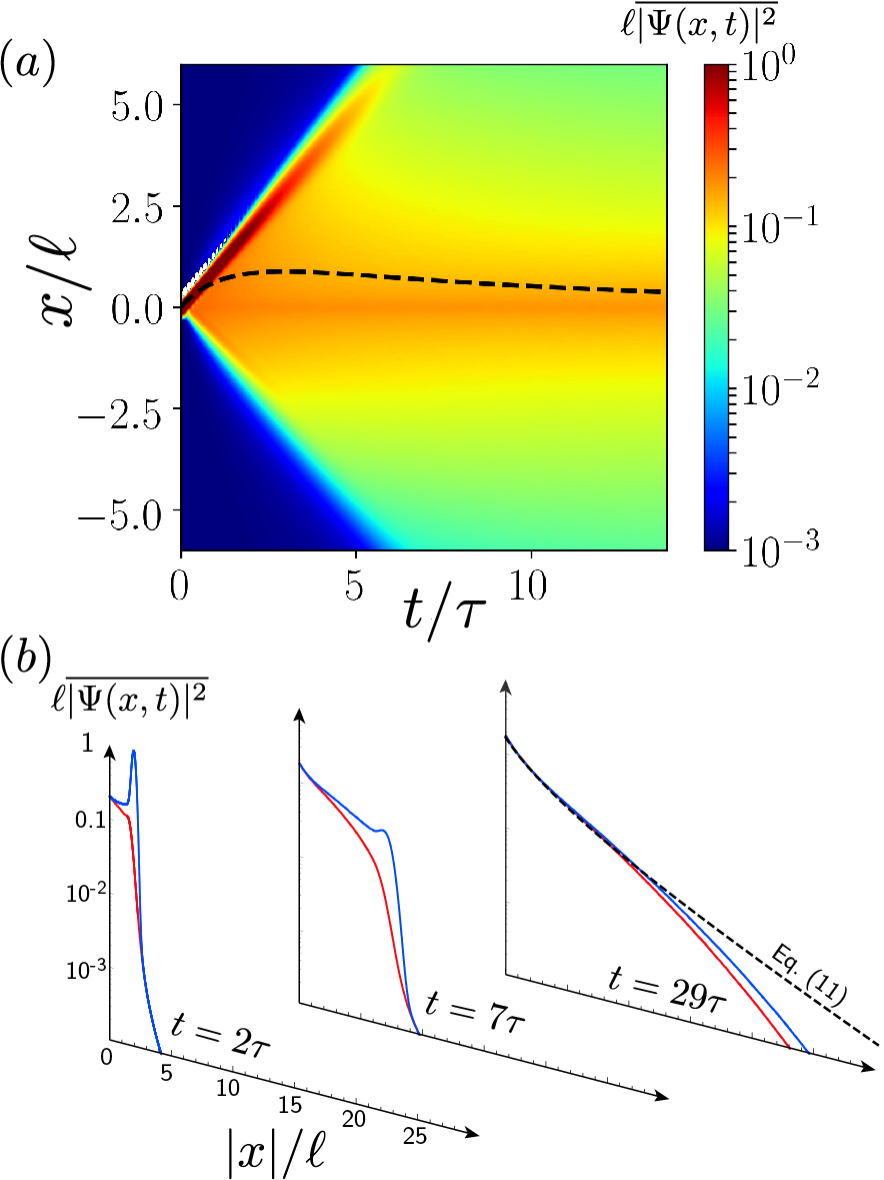}
	\end{center}
	\caption{
	(a) Numerical density plot of the average density profile as a function of space and time, showing localization at long time. The dashed black curve is the position of the center of mass $\langle x(t)\rangle$ which first increases and eventually comes back to its initial value.
	(b) Average density profile at three successive times.
		The solid upper blue and lower red curves are the $x>0$ and $x<0$ components of the profile, respectively. The long-time limit of the profile, Eq. \eqref{eq-Gogolin-profile}, is shown as a dashed black curve.	
		}
	\label{fig-profiles}
\end{figure}

In order to clarify which specific behavior of the spatial distribution $\overline{|\Psi(x,t)|^2}$ actually gives rise to the QB phenomenon, 
we show in Fig. \ref{fig-profiles}(a) a numerical density plot of the average density profile $\overline{|\Psi(x,t)|^2}$ as a function of space and time, indicating on the top the position of the center of mass. Figure \ref{fig-profiles}(b) also shows the $x\!>\!0$ (blue curve) and $x\!<\!0$ (red curve) components of $\overline{|\Psi(x,t)|^2}$ at three successive times. The profiles display a ballistic peak responsible for the increase of $\langle x(t)\rangle $ at short times. 
After this peak has been attenuated, the profile re-symmetrizes itself around $x\!=\!0$, which gives rise to the QB effect.
As discussed above, at long time the distribution converges toward a symmetric one, Eq. (\ref{eq-inf-time-prof-mode-exp}), the so-called Gogolin density profile \cite{Gogolin82,footnote1}:
\begin{equation}
\overline{\vert \Psi (x,\infty) \vert^2} = \int_0^{\infty}  \frac{\text{d} \eta \pi^2}{32 \ell} \frac{\eta \left(1+\eta^2\right)^2 \sinh (\pi  \eta) e^{-\left(1+\eta^2\right) \left| x\right| /8 \ell}}{[1+\cosh (\pi \eta)]^2},
 \label{eq-Gogolin-profile}
\end{equation}
which is shown in Fig. \ref{fig-profiles}(b) for comparison. Note that
although we start from a rather narrow wave packet with $\sigma\!<\!\ell$ in our simulations, the QB phenomenon is present as well when $\sigma\!>\!\ell$. It is however less dramatic because the forth and back motion of the CoM then has an amplitude smaller than the wave packet size.

A natural question is  whether the QB effect also exists in higher dimension. The  Berezinskii technique is specific to 1D systems and cannot be used, but the general spectral argument discussed above - based on Eqs.~(\ref{eq-exp-psi-eigenstates},\ref{eq-inf-time-prof-mode-exp}) - suggests that the QB effect should occur if the eigenstates are Anderson localized. In contrast, if the eigenstates are extended, no minimum energy difference exists for the non-diagonal parts of Eq.~(\ref{eq-exp-psi-eigenstates}), so that the argument does not apply. This is especially the case in diffusive systems, where quantum interference effects are very small.
In order to test this scenario, we have performed numerical experiments using the three-dimensional (3D) Anderson model, that is a 3D cubic lattice with Hamiltonian
\begin{equation}
H = \sum_i{\epsilon_i |i\rangle \langle i |} - \sum_{<i,j>}{|i \rangle \langle j |} 
\end{equation} 
where $i,j$ denote sites of the 3D lattice, $\epsilon_i$ are independent random energies uniformly distributed in $[-W/2,W/2]$ and the second sum involves only neighboring pairs of sites, with the hopping amplitude taken as the energy unit. The phase diagram of this model is well known~\cite{Kroha90}. In particular, states with energy near $E=0$ undergo a transition - known as the Anderson transition - between extended states at low disorder $W$ and localized states at strong disorder. The ``mobility edge'' separating the two regimes is given by $W_c\approx 16.54$~\cite{Slevin14}. 

We first prepare a moving Gaussian wave packet at the center of the system, 
$\psi(\br)\!\propto \!\exp(-\br^2/2\sigma^2\!+\!i\bk_0.\br),$ with $\sigma\!=\!10$ lattice sites and an initial momentum $\bk_0\!=\!(\pi/2,\pi/2,\pi/2)$ in units of the inverse of the lattice step. We numerically propagate this wave packet in the presence of disorder and monitor the CoM position $\langle \psi(t)|\br|\psi(t)\rangle$ \cite{Filter}. The system size is chosen sufficiently large for the wavefunction to be negligibly small at the system edges at the longest times considered, so that the CoM position is computed for a virtually infinite system. 
$\langle \psi(t)|\br|\psi(t)\rangle$ is averaged over many (typically few thousands) realizations of the disorder. Because the system is statistically invariant by permutation of the 3 axes of the cubic lattice, the averages $\langle x(t)\rangle, \langle y(t)\rangle, \langle z(t)\rangle$ are equal. In the following, we use $\langle x(t)\rangle$ as a shorthand notation for an additional averaging along the 3 directions.

\begin{figure}
	\begin{center}
		\includegraphics[width=\linewidth]{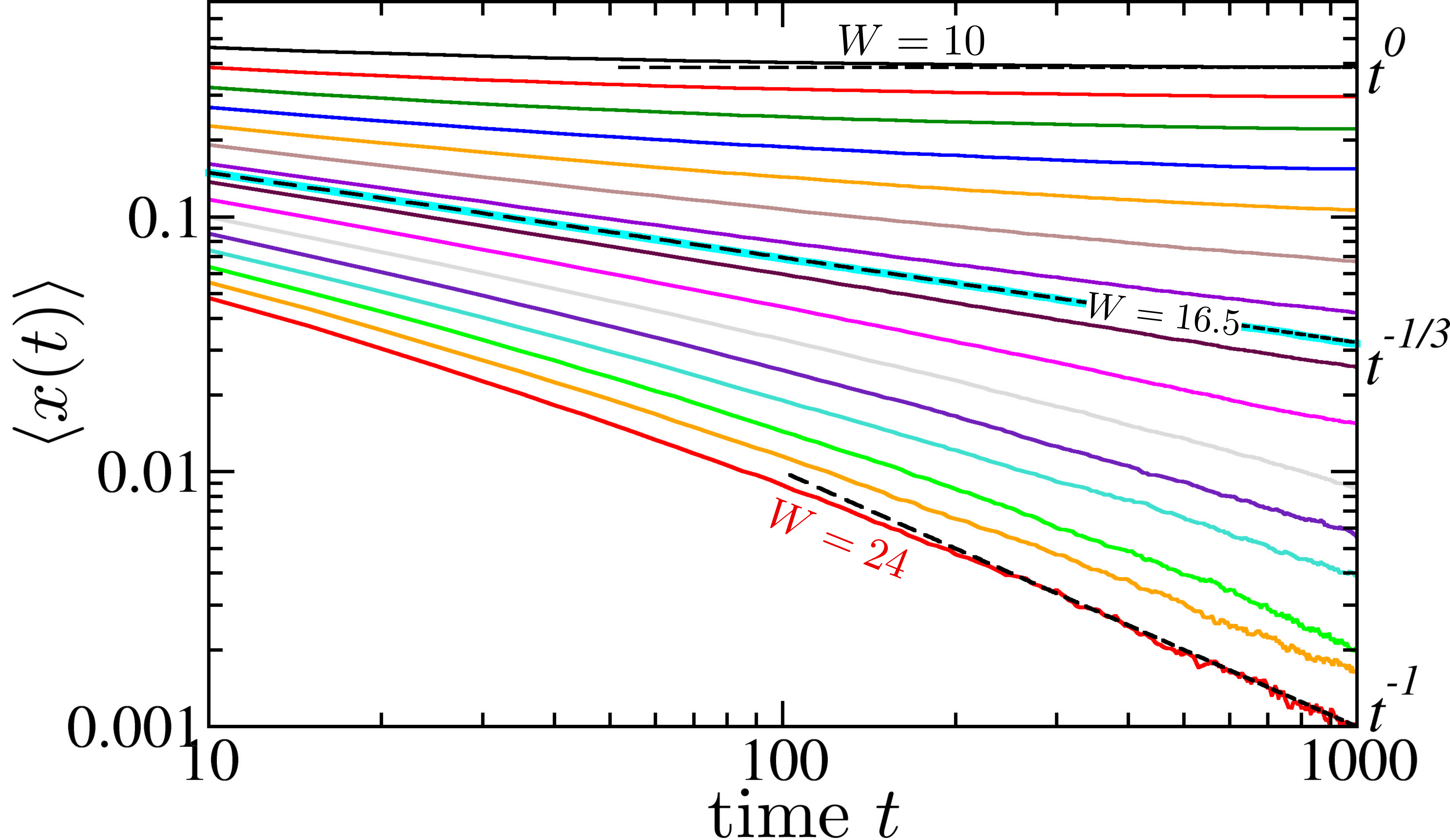}
	\end{center}
	\caption{
	Average center of mass versus time in three dimensions. Data are for the 3D Anderson model, at various values of disorder $W,$ from 10 to 24 (with step 1). Here position is in units of the lattice spacing and time in units of the inverse of the hopping matrix element. For $W\!<\!W_c\approx 16.5,$  $\langle x(t)\rangle$ saturates  to a finite value. In contrast, for $W\!>\!W_c$, a full quantum boomerang effect is clearly visible, with the center of mass returning to its initial position, with an asymptotic behavior $\propto\!1/t.$ For $W\!=\!W_c$ (thick cyan line), the critical behavior is such that $\langle x(t)\rangle\!\propto\!t^{-1/3}.$ The dashed lines indicate the asymptotic dependences at long time for the extended, critical and localized regimes.
	These results show that the quantum boomerang effect is a clear-cut signature of Anderson localization.}
	\label{fig-dispersion-3d}
\end{figure}
Fig.~\ref{fig-dispersion-3d} shows the temporal evolution of  $\langle x(t)\rangle$ for various disorder strengths $W$ up to $t=1000$ (with time in units of the inverse of the hopping amplitude). In all cases, the short time dynamics - not shown in the figure - is ballistic, with  $\langle x(t)\rangle$ increasing with $t,$ but soon the wave packet performs a U-turn so that $\langle x(t)\rangle$ decreases.
For $W\!<\!W_c,$ $\langle x(t)\rangle$ tends to a \emph{finite non-zero} value at long time, indicating a breakdown of the QB effect. In contrast, for $W\!>\!W_c,$ $\langle x(t)\rangle$ tends to zero, a manifestation of the QB. The long-time behavior is approximately $\propto \!1/t,$ that is slower than in 1D, Eq~(\ref{results_mean_x}). The numerical results are not accurate enough to assess whether there is e.g an additional logarithmic dependence. We performed similar calculations in dimension 2, where Anderson localization is the generic scenario and observed a similar $1/t$ asymptotic behavior (data not shown). 

\begin{figure}
\begin{center}
		\includegraphics[width=0.9\linewidth]{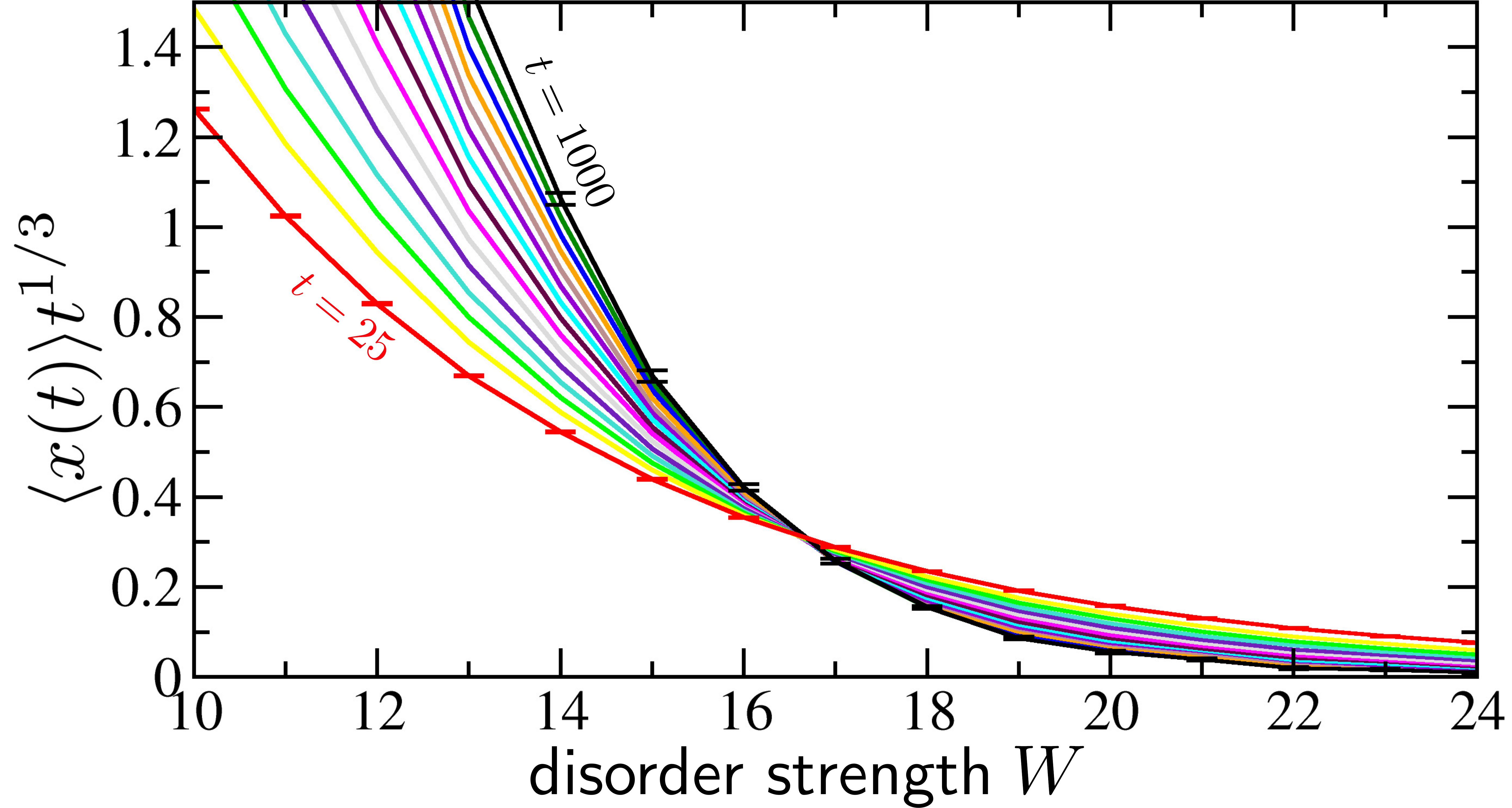}
\end{center}
 \caption{
 Quantum boomerang effect in the vicinity of the Anderson transition (same units as in Fig. \ref{fig-dispersion-3d}).
Solid curves show the quantity $\langle x(t)\rangle t^{1/3}$ versus the disorder strength $W,$ for increasing times, from $t\!=\!25$, flattest curve in red, to $t\!=\!1000$, top black curve (times are equally spaced in $t^{1/3}$).
 All curves intersect at $W_c\!\approx\!16.55\pm 0.03,$ making it possible to accurately pinpoint the critical point of the Anderson transition.
}
 \label{fig-scaling-3d}
\end{figure}

At the critical point of the Anderson transition, $W\!=\!W_c,$ the QB effect is present, albeit with a slower decay, very accurately described by a $t^{-1/3}$ law. This law is reminiscent of the anomalous diffusion at the critical point with the size of a wave packet increasing like $t^{1/3},$ predicted theoretically~\cite{Shapiro82,Kroha90,Mueller16} and experimentally observed on the atomic kicked rotor~\cite{Lemarie10}. Indeed, by extrapolating Eq. (\ref{eq-relation_x_xx}) to 3D and using $\langle x^2\rangle \propto t^{2/3}$ one readily obtains $\langle x\rangle\propto t^{-1/3}$. The same argument explains why $\langle x\rangle$ is constant above the critical point, as $\langle x^2\rangle\propto D t$ in this regime, with a diffusion coefficient $D$ getting smaller and smaller as one approaches the Anderson transition. These different scaling law can be used
 to pinpoint precisely the critical point. To this aim, we show in Fig.~\ref{fig-scaling-3d} the quantity  $\langle x(t)\rangle t^{1/3}$ vs. $W$ for increasingly long times.
As expected, all curves cross near $W\!=\!W_c.$ From our numerical data, we observe the crossing of long-time curves at $W\!=\!16.55\pm0.03,$ in excellent agreement with the most accurate published value $W_c=16.543\pm0.002$~\cite{Slevin14,Slevin18}. 

To summarize, we have demonstrated that wave packets launched with some initial velocity in disordered systems quantum-mechanically return to their initial position, an effect absent  in the classical limit. An exact analytical treatment for 1D weakly disordered systems is in excellent agreement with numerical results. In dimension 3, the quantum boomerang effect exists only when the eigenstates are Anderson localized; In the diffusive regime, the center of mass displays only a partial retroreflection and does not end at the initial position.  We thus expect the quantum boomerang to be very general for 
disordered systems displaying Anderson localization. Based on the discussion around Eqs.~(\ref{eq-exp-psi-eigenstates},\ref{eq-inf-time-prof-mode-exp}), it only requires time-reversal symmetry, statistical invariance by translation and parity of the disorder. Whether it persists when these symmetries are broken or  for many-body localized systems will be the subject of further studies. 
From an experimental point of view, this signature of Anderson localization could be observed in experiments on cold atomic gases, where wave-packet spatial distributions are customarily imaged. In optics, the retroreflection could be probed as well in transversally disordered photorefractive crystals or optical fibers where the coordinate of the optical axis plays the role of time: by illuminating such systems with a spatially narrow beam, one should observe a quantum boomerang of the beam center of mass in the transmitted, near-field intensity distribution \cite{Schwartz07, Boguslawski17}.

We thank Christian Miniatura and Kean-Loon Lee for discussions at the early stages of this work, and Bogdan Damski for suggesting the name ``quantum boomerang''.



\begin{thebibliography}{99}
\bibitem{footnote4}
This can be justified more rigorously by using the phase space Wigner representation $\overline{W(x,p,t)}$ of the wave packet. 
In the weak-disorder approximation, $\overline{W(x,p,t)}$ is non vanishing only near $p=\pm\hbar k_0,$ 
so that it can be split in two parts $\overline{W(x,p,t)}\approx n_+(x,t) \delta(p-\hbar k_0)+n_-(x,t) \delta(p+\hbar k_0)$.
\bibitem{footnote5}
Although the calculations in \cite{Nakhmedov87,Berezinsky73} are correct, there are occasionally factors 2 missing in some equations.
\end{thebibliography}
\end{document}